\documentclass[10pt,prd,superscriptaddress,preprintnumbers,amsmath,amssymb,nofootinbib]{revtex4}
\usepackage{epsfig}  
\usepackage{graphicx}
\usepackage{hyperref}
\usepackage{color}
\usepackage{float}
\usepackage{amsfonts}
\usepackage{amsmath}
\usepackage{slashed}
\usepackage{placeins}

\large

\begin{document} 
\title{New physics effects in purely leptonic $B^*_s$ decays}

\author{Suman Kumbhakar}
\email{suman@phy.iitb.ac.in}
\affiliation{Indian Institute of Technology Bombay, Mumbai 400076, India}

\author{Jyoti Saini} 
\email{saini.1@iitj.ac.in }
\affiliation{Indian Institute of Technology Jodhpur, Jodhpur 342037, India}

\date{\today} 

\preprint{}

\begin{abstract}

Recently several measurements in the neutral current  sector $b\rightarrow s l^+l^-$ ($l=e$ or $\mu$) as well as in the charged current sector $b \rightarrow c \tau \bar{\nu}$ show significant deviations from their Standard Model predictions. It has been shown that two different new physics solutions can explain all the anomalies in $b\rightarrow s l^+l^-$ sector. Both these solutions are in the form of linear combinations  of the two operators $(\bar{s}\gamma^{\alpha}P_Lb)(\bar{\mu}\gamma_{\alpha}\mu)$ and $(\bar{s}\gamma^{\alpha}P_Lb)(\bar{\mu}\gamma_{\alpha}\gamma_5\mu)$. We show that the longitudinal polarization asymmetry of the muons in $B^*_s\rightarrow \mu^+\,\mu^-$ decay is a good discriminant between the two solutions if it can be measured to a precision of $10\%$, provided the new physics Wilson coefficients are real. If they are complex, the theoretical uncertainties in this asymmetry are too large to provide effective discrimination. We also investigate the potential impact of $b \rightarrow c \tau \bar{\nu}$ anomalies on $b \rightarrow s \tau^+ \tau^-$ transitions. We consider a model where the new phyics contributions to these two transitions are strongly correlated. We find that the branching ratio of $B^*_s\rightarrow \tau^+\,\tau^-$ can be enhanced by three orders of magnitude.

\end{abstract}

\maketitle 

%%%%%%%%%%%%%%%%%%%%%%%%%%%%
\section{Introduction} 
%%%%%%%%%%%%%%%%%%%%%%%%%%%%
In the past few years, a number of anomalies have been observed in the decays of $B$ mesons. They occur both in the charged current (CC) transition $b\rightarrow c \tau\bar{\nu}$ and in the flavor changing neutral current (FCNC) transitions $b\rightarrow sl^+l^-$ ($l=e$ or $\mu$). In the Standard Model (SM), the above CC transition occurs at tree level whereas the FCNC transitions occur only at loop level. The discrepancies, between the measured values and the SM predictions, vary for different observables. 

First, we discuss the anomalies in $b\rightarrow sl^+l^-$ transitions. They are
\begin{enumerate}
\item In the decay $B\rightarrow K^* \mu^+ \mu^-$, some of the angular observables \cite{Aaij2013,Aaij2016,Abdesselam} are found to be in disagreement with their respective SM predictions \cite{Descotes}. The main discrepancy is in the angular observable $P^{'}_5$, which is at the level of $4~\sigma$. 		
\item  The branching ratio of $B_s\rightarrow \phi \mu^+ \mu^-$ and the corresponding angular observables also differ from their SM predictions~\cite{Aaij2013JHEP,Aaij20171509} at $3.5~\sigma$ level.
\item The SM predicts the ratio $R_K \equiv \Gamma(B^+ \rightarrow K^+ \mu^+ \mu^-)/\Gamma(B^+ \rightarrow K^+ e^+ e^-)\simeq 1 $. LHCb experiment measured this ratio in the $q^2$ ($q^2=(p_B-p_K)^2$) range $1.0 \leq q^2 \leq 6.0$ GeV$^2$~\cite{Aaij2014}. The measured value $0.745^{+0.090}_{-0.074}(stat.)\pm 0.036(syst.)$ deviates from the SM prediction by $2.6~\sigma$~\cite{Hiller,Bordone}. 
\item LHCb experiment also measured the ratio $R_{K^*} \equiv \Gamma(B^0 \rightarrow K^{*0} \mu^+ \mu^-)/\Gamma(B^0 \rightarrow K^{*0} e^+ e^-)$ in two different $q^2$ ranges, $(0.045 \leq q^2 \leq 1.1$ GeV$^2)$ (low $q^2$) and $(1.1 \leq q^2 \leq 6.0$ GeV$^2)$ (central $q^2$). The SM predicts this ratio to be $\simeq$ 1 for all $q^2$~\cite{Hiller,Bordone}. The measured values are $0.660^{+0.110}_{-0.070}(stat.)\pm 0.024(syst.)$ for low $q^2$ and $0.685^{+0.113}_{-0.069}(stat.)\pm 0.047(syst.)$ for central $q^2$~\cite{Aaij2017}. These differ from the SM prediction by $2.2-2.4 ~\sigma$ and $2.4-2.5~\sigma$ respectively. 
\end{enumerate}

The anomalies in $R_K$ and $R_{K^*}$, which are an indication of violation of lepton flavor universality (LFU) in the neutral current decays of b quark, can be explained by new physics (NP) in either $b\rightarrow se^+e^-$ or $b\rightarrow s\mu^+\mu^-$ or both whereas the first two anomalies require NP in $b\rightarrow s\mu^+\mu^-$. Two kinds of NP amplitudes in $b\rightarrow se^+e^-$ transitions can account for the $R_K$ and $R_{K^*}$ anomalies. These are 
\begin{itemize}
\item vector and/or axial-vector amplitudes which will have constructive interference with the SM amplitude. The magnitude of such amplitude should be about $10\%$ of the SM amplitude.
\item scalar, pseudoscalar or tensor amplitudes which do not interfere with the SM amplitude. A discussion of the most general NP contribution to $b\rightarrow se^+e^-$ is beyond the scope of this paper.
\end{itemize}
In this work, we will consider NP amplitudes only in $b\rightarrow s\mu^+\mu^-$ transition, because they can explain all four anomalies in the FCNC decays of the $B$ mesons. These amplitudes \textbf{must have destructive interference} with the SM amplitude so that the resulting values of $R_K$ and $R_{K^*}$ will be less than 1. That is, these NP amplitudes are constrained to be vector and/or axial-vector amplitudes. Several groups \cite{Capdevila:2017bsm,Altmannshofer:prd96, Amico:jhep1709,Hiller:prd96,Grinstein:prd093006,Ciuchini:epjc77,Celis:prd035026,AKAPRD96,AKAPRD96:015034,Arbey:2018ics} have performed global fits to identify the Lorentz structure of the NP operators and to determine their Wilson coefficients (WCs) which can account for all the $b\rightarrow s \mu^+ \mu^-$ anomalies. There are two distinct solutions, one with the operator of the form $(\bar{s}\gamma^{\alpha}P_Lb)(\bar{\mu}\gamma_{\alpha}\mu)$ and the other whose operator is a linear combination of $(\bar{s}\gamma^{\alpha}P_Lb)(\bar{\mu}\gamma_{\alpha}\mu)$ and $(\bar{s}\gamma^{\alpha}P_Lb)(\bar{\mu}\gamma_{\alpha}\gamma_5\mu)$~\cite{AKAPRD96}. These results satisfy the requirement that only vector ($V$) and/or axial-vector ($A$) NP operators are allowed.

It is interesting to look for new observables in the $b \rightarrow s \mu^+ \mu^-$ sector in order to (a) find additional evidence for the existence of NP and (b) to discriminate between the two NP solutions. These observables may be related to the observed decay modes or may be associated with the decay modes yet to be observed such as $B_s\rightarrow l^+l^-\gamma$~\cite{Abbas}.
 
The branching ratio of $B_s^*$ meson to di-muons is one such observable which is yet to be measured. In the SM, this decay mode is not subject to helicity suppression~\cite{Grinstein1509}, unlike $B_s\rightarrow \mu^+\mu^-$~\cite{Fleischer:2017ltw}. Further, it is sensitive to NP operators containing both $V$ and $A$ currents of leptons whereas $B_s\rightarrow \mu^+\mu^-$ is sensitive only to the latter. A model independent analysis of this decay was performed in ref.~\cite{Kumar:2017xgl} to identify the NP operators which can lead to a large enhancement of its branching ratio. It was found that such an enhancement is not possible due to the constraints from the present $b\rightarrow s\mu^+\mu^-$ data. It would be desirable to construct a new observable related to this decay mode to see whether such an observable has the potential to discriminate between the two existing NP solutions in $b\rightarrow s \mu^+ \mu^-$ transition. 
  
  In this work, we consider the longitudinal polarization asymmetry of muon in $B^*_s \rightarrow \mu^+ \mu^-$ decay, $\mathcal{A}_{LP}(\mu)$. This asymmetry is theoretically clean because it has a very mild dependence on the decay constants unlike the branching ratio. We first calculate the SM prediction of  $\mathcal{A}_{LP}(\mu)$ and then study its sensitivity to the two NP solutions. 

As mentioned above, there are additional discrepancies in the CC decays of $B$ mesons. Such decays are driven by $b\rightarrow c \tau \bar{\nu}$ transition, which occurs at tree level in the SM. These discrepancies, which are listed below, are an indication of LFU violation in the charged current decays of $b$ quark. 
    \begin{enumerate}
     \item The current world averge of the ratio $R_{D} = \mathcal{B}(B\rightarrow D\,\tau\,\bar{\nu})/\mathcal{B}(B \rightarrow D\{e/ \mu\}\, \bar{\nu})$, measured by BaBar and Belle, deviates $2.3\sigma$ from the SM prediction~\cite{HFAG:2017avg}.
      \item There is a series of measurements of the ratio $R_{D^*} = \mathcal{B}(B\rightarrow D^*\,\tau\,\bar{\nu})/\mathcal{B}(B \rightarrow D^*\{e/ \mu\}\, \bar{\nu})$ by BaBar, Belle and LHCb experiments. Recent world average of $R_{D^*}$ shows a discrepancy with respect to the SM prediction at a level of $3.4\sigma$.  Including the measurement correlation between $R_D$ and $R_{D^*}$, the current experimental world averages of $R_{D^{(*)}}$ show a $\sim$ $4\sigma$ deviation from the SM predictions~\cite{HFAG:2017avg}.
      \item The measured value of $R_{J/ \psi} = \mathcal{B}(B\rightarrow J/ \psi\,\tau\,\bar{\nu})/\mathcal{B}(B \rightarrow J/ \psi \mu\, \bar{\nu})$ by LHCb collaboration, is $1.7 \sigma$ away from its SM prediction~\cite{Aaij:2017tyk}. 
    \end{enumerate}
The NP operators which can account for $R_{D^{(*)}}$ anomaly are identified in ref~\cite{Ligeti}. In ref.~\cite{Alok:2017qsi} it was shown that there are only four independent NP solutions which can explain the present data in the  $b \rightarrow c \tau \bar{\nu}$ sector. Methods to discriminate between these NP solutions were suggested in ref.~\cite{Alok:2018uft}. The NP WCs of these solutions are about $10\%$ of the SM values. Since this transition occurs at tree level in the SM, it is very likely that the NP operators also occur at tree level. In the SM, the relation between the interaction eigenstates and mass eigenstates leads to the cancellation of FCNCs at tree level through GIM mechanism. However the relation between the interaction eigenstates of NP and the mass eigenstates need not be the same as that in the SM. In such a situation, the NP will lead to tree level neutral current $b\rightarrow s l^+l^-$ transitions. In ref.~\cite{Capdevila:2017iqn}, a model is constructed where the tree level FCNC terms due to NP are significant for $b\rightarrow s\,\tau^+\,\tau^-$ but are suppressed for $b\rightarrow sl^+l^-$ where $l=e$ or $l=\mu$. The branching ratios for the decay modes such as $B\rightarrow K^{(*)}\tau^+\tau^-$, $B_s\rightarrow \tau^+\tau^-$ and $B_s\rightarrow \phi\tau^+\tau^-$ will have a large enhancement in this model~\cite{Capdevila:2017iqn}. In this work we study the effect of this NP on the branching ratio of $B_s^* \rightarrow \tau^+ \tau^-$ and the $\tau$ polarization asymmetry $\mathcal{A}_{LP}(\tau)$.

This paper is organized as follows. In section II, we obtain the theoretical expressions for the longitudinal polarization asymmetry of the final state leptons in $B^*_s \rightarrow l^+\, l^-$ decays, where $l=e,\mu$ or $\tau$. This is done for the SM and for the case of NP $V$ and $A$ operators. In section III, we obtain predictions of  $\mathcal{A}_{LP}(\mu)$ in both the SM and the two NP solutions which explain all $ b \rightarrow s \mu^+ \mu^-$ anomalies. In the same section we study the impact of tree level NP of ref.~\cite{Capdevila:2017iqn} on the branching ratio of $B_s^* \rightarrow \tau^+ \tau^-$ and $\mathcal{A}_{LP}(\tau)$. Finally in the section IV, we present our conclusions.

%%%%%%%%%%%%%%%%%%%%%%%%%%%%
\section{Calculation of Longitudinal Polarization Asymmetry for $B_s^* \rightarrow l^+ l^-$ decay }
%%%%%%%%%%%%%%%%%%%%%%%%%%%%
\subsection{Longitudinal Polarization Asymmetry in the SM}
The pure leptonic decay  $B_s^* \rightarrow l^+\, l^-$ is induced by the quark level transition $ b\rightarrow s l^+ l^-$. In the SM the corresponding effective Hamiltonian is
\begin{align} \nonumber
\mathcal{H}_{SM} &= − \frac{4 G_F}{\sqrt{2} \pi} V_{ts}^* V_{tb} \bigg[ \sum_{i=1}^{6} C_i(\mu) O_i(\mu) + C_7 \frac{e}{16 \pi^2} [\overline{s} \sigma_{\mu \nu}(m_s P_L  + \\ \nonumber &
m_b P_R)b]F^{\mu \nu} + C_9 \frac{\alpha_{em}}{4 \pi}(\overline{s} \gamma^{\mu} P_L b)(\overline{l} \gamma_{\mu} l) + C_{10} \frac{\alpha_{em}}{4 \pi} \\  &
 (\overline{s} \gamma^{\mu} P_L b)(\overline{l} \gamma_{\mu} \gamma_5 l) \bigg],
\end{align}
where $G_F$ is the Fermi constant, $V_{ts}$ and $V_{tb}$ are the Cabibbo-Kobayashi-Maskawa (CKM) matrix elements and $P_{L,R} = (1 \mp \gamma^{5})/2$ are the projection operators. The effect of the operators $O_i,\,i=1-6,8 $ can be embedded in the redefined effective Wilson coefficients as $C_7(\mu)\rightarrow C^{eff}_7(\mu,q^2)$ and $C_9(\mu)\rightarrow C^{eff}_9(\mu,q^2)$.

The $B_s^* \rightarrow l^+\, l^-$  amplitude can be parameterized in terms of the following form factors~\cite{Grinstein1509}
\begin{align}\nonumber
\langle 0 \vert\overline{s}\gamma^{\mu} b \vert B_s^{*}(p_{B_s^*},\epsilon) \rangle &= f_{B_s^{*}}m_{B_s^*} \epsilon^{\mu}, \\ \nonumber
\langle 0 \vert\overline{s}\sigma^{\mu \nu} b \vert B_s^{*}(p_{B_s^*},\epsilon) \rangle &= - i f^T_{B_s^{*}}(p^{\mu}_{B_s^{*}} \epsilon^{\nu} - \epsilon^{\mu}  p^{\nu}_{B_s^{*}} ),  \\
\langle0 \vert\overline{s}\gamma^{\mu} \gamma_{5} b \vert B_s^{*}(p_{B_s^*},\epsilon) \rangle &= 0, 
\end{align}
where $\epsilon^{\mu}$ is the polarization vector of the $B^*_s$ meson and $f_{B^*_s}$ and $f_{B^*_s}^T$ are the decay constants of $B^*_s$ meson. In the heavy quark limit they are related to $f_{B_s}$, the decay constant of $B_s$ meson, as
\begin{align}\nonumber
f_{B_s^*} &= f_{B_s} \left[ 1- \frac{2 \alpha_s}{3 \pi}\right], \\ 
f_{B_s^*}^T &= f_{B_s} \left[ 1 + \frac{2 \alpha_s}{3 \pi}\left(\log(\frac{m_b}{\mu}) - 1\right)\right].
\end{align}
The decay constant $f_{B_s}$ is defined through the relation
$$\langle 0 \vert\overline{s}\gamma^{\mu} \gamma_{5} b \vert B_s(p_{B_s}) \rangle = - i f_{B_s}p_{B_s}^{\mu}.$$
The SM amplitude for $B^*_s \rightarrow l^+\, l^-$ decay is given by~
\begin{widetext}
\begin{align}
\mathcal{M}_{SM} &= -\frac{\alpha_{em}  G_F }{2 \sqrt{2}\pi }f_{B^*_s}V_{ts}^{*} V_{tb}\,m_{B_s^*} \,\epsilon^{\mu} \left[   \left(C_9^{eff} + \frac{2 m_b f_{B_s^*}^T}{m_{B_s^*} f_{B_s^*}} C_7^{eff}\right)\left(\overline{l} \gamma_{\mu} l\right) + C_{10}\left(\overline{l} \gamma_{\mu} \gamma_5 l\right)  \right],
\end{align}
\end{widetext}
and the decay rate is found to be
\begin{widetext}
\begin{equation}
\Gamma_{SM} = \frac{\alpha^2_{em}G^2_Ff^2_{B^*_s}m^3_{B^*_s}}{96\pi^3}\vert V_{ts}V^*_{tb}\vert^2\sqrt{1-4m_l^2/m^2_{B^*_s}}\left[\left(1+\frac{2m^2_l}{m^2_{B^*_s}}\right)\left\vert C_9^{eff}+\frac{2 m_b f_{B_s^*}^T}{m_{B_s^*} f_{B_s^*}}C_7^{eff}\right\vert^2+\left(1-\frac{4m^2_l}{m^2_{B^*_s}}\right)\vert C_{10}\vert^2\right].
\label{br}
\end{equation}
\end{widetext}

We define the longitudinal polarization asymmetry for the final state leptons in $B_s^* \rightarrow l^+ l^-$ decay. The unit longitudinal polarization four-vector in the rest frame of the lepton ($l^{+}$ or $l^-$) is defined as
\begin{align}
\overline{s}_{l^{\pm}}^{\alpha}= \left(0, \pm\frac{\overrightarrow{p_{l}}}{|\overrightarrow{p_{l}}|}\right).
\end{align} 
In the dilepton rest frame  (which is also the rest frame of $B_s^* $ meson), these unit polarization vectors become
\begin{align}
s_{l^{\pm}}^{\alpha}= \left(\frac{|\overrightarrow{p_{l}}|}{m_{\l}}, \pm \frac{E_{l}}{m_{l}} \frac{\overrightarrow{p_{l}}}{|\overrightarrow{p_{l}}|}\right),
\end{align}
where $E_{l}$, $\overrightarrow{p_{l}}$ and $m_{l}$ are the energy, momentum and mass of the lepton ($l^{+}$ or $l^-$) respectively. We can define two longitudinal polarization asymmetries, $\mathcal{A}_{LP}^{+}$ for $l^+$ and $\mathcal{A}_{LP}^{-}$ for $l^-$, in the decay $B_s^* \rightarrow l^+\, l^-$ as~\cite{Handoko:prd65,Alok:2008hh,AKA121}
\begin{widetext}
\begin{equation}
\mathcal{A}_{LP}^{\pm}=\frac{[\Gamma( s_{l^-}, s_{l^+}) + \Gamma(\mp s_{l^-}, \pm s_{l^+})]-[\Gamma(\pm s_{l^-}, \mp s_{l^+}) + \Gamma(-s_{l^-},-s_{l^+})]}{[\Gamma( s_{l^-}, s_{l^+}) + \Gamma(\mp s_{l^-}, \pm s_{l^+})]+[\Gamma(\pm s_{l^-}, \mp s_{l^+}) + \Gamma(-s_{l^-},-s_{l^+})]}.
\label{alp}
\end{equation}
\end{widetext}
If the two spin projections, $s_{l^-}$ and $s_{l^+}$ are the same, the decay rate is given by
\begin{widetext}
\begin{eqnarray}
\Gamma (\pm s_{l^-}, \pm s_{l^+})&=& \mathcal{N}\left[\frac{4m^2_l}{3}\vert \mathcal{C}\vert^2 +\frac{\mathcal{C}C_{10}^*}{6m_lm_{B^*_s}}\left\lbrace i\sqrt{m^2_{B^*_s}-4m^2_l}\left(\varepsilon_{\alpha\beta\gamma\nu}p^{\alpha}_{l^-}p^{\beta}_{B^*_s}p^{\gamma}_{l^+}s^{\nu}_{l^-} +\varepsilon_{\alpha\beta\gamma\sigma}p^{\alpha}_{l^-}p^{\beta}_{B^*_s}p^{\gamma}_{l^+}s^{\sigma}_{l^+}\right)\right.\right.\nonumber\\
& & \left.  + i m_lm_{B^*_s}\left(\varepsilon_{\alpha\beta\nu\sigma}p^{\alpha}_{l^-}p^{\beta}_{B^*_s}s^{\nu}_{l^-}s^{\sigma}_{l^+}-\varepsilon_{\beta\gamma\nu\sigma} p^{\beta}_{B^*_s}p^{\gamma}_{l^+}s^{\nu}_{l^-}s^{\sigma}_{l^+}\right)\right\rbrace +\frac{\mathcal{C}^*C_{10}}{6m_lm_{B^*_s}}\left\lbrace -i\sqrt{m^2_{B^*_s}-4m^2_l} \right.\nonumber\\
 & &\left.\left. \left(\varepsilon_{\alpha\beta\gamma\nu}p^{\alpha}_{l^-}p^{\beta}_{B^*_s}p^{\gamma}_{l^+}s^{\nu}_{l^-}+\varepsilon_{\alpha\beta\gamma\sigma}p^{\alpha}_{l^-}p^{\beta}_{B^*_s}p^{\gamma}_{l^+}s^{\sigma}_{l^+}\right) -im_lm_{B^*_s}\left(\varepsilon_{\alpha\beta\nu\sigma}p^{\alpha}_{l^-}p^{\beta}_{B^*_s}s^{\nu}_{l^-}s^{\sigma}_{l^+}-\varepsilon_{\beta\gamma\nu\sigma}p^{\beta}_{B^*_s}p^{\gamma}_{l^+}s^{\nu}_{l^-}s^{\sigma}_{l^+}\right)\right\rbrace\right],
\label{gama1}
\end{eqnarray}
\end{widetext}
For opposite spin projections of $s_{l^-}$ and $s_{l^+}$ we have
\begin{widetext}
\begin{eqnarray}
\Gamma (\mp s_{l^-},\pm s_{l^+}) &=& \mathcal{N}\left[\frac{2m^2_{B^*_s}}{3}\vert \mathcal{C}\vert^2 +\frac{\mathcal{C}C^*_{10}}{6m_lm_{B^*_s}}\left\lbrace m_lm_{B^*_s}\left(-i\varepsilon_{\alpha\beta\nu\sigma}p^{\alpha}_{l^-}p^{\beta}_{B^*_s}s^{\nu}_{l^-}s^{\sigma}_{l^+}+i\varepsilon_{\beta\gamma\nu\sigma}p^{\beta}_{B^*_s}p^{\gamma}_{l^+}s^{\nu}_{l^-}s^{\sigma}_{l^+}\right.\right.\right.\nonumber\\
& & \left.\left. \mp 4m_{B^*_s}\sqrt{m^2_{B^*_s}-4m^2_l}\right) -i\sqrt{m^2_{B^*_s}-4m^2_l}\left(\varepsilon_{\alpha\beta\gamma\nu}p^{\alpha}_{l^-}p^{\beta}_{B^*_s}p^{\gamma}_{l^+}s^{\nu}_{l^-}+\varepsilon_{\alpha\beta\gamma\sigma}p^{\alpha}_{l^-}p^{\beta}_{B^*_s}p^{\gamma}_{l^+}s^{\sigma}_{l^+}\right)\right\rbrace \nonumber\\
& & +\frac{\mathcal{C}^*C_{10}}{6m_lm_{B^*_s}}\left\lbrace i\sqrt{m^2_{B^*_s}-4m^2_l}\left(\varepsilon_{\alpha\beta\gamma\nu}p^{\alpha}_{l^-}p^{\beta}_{B^*_s}p^{\gamma}_{l^+}s^{\nu}_{l^-}+\varepsilon_{\alpha\beta\gamma\sigma}p^{\alpha}_{l^-}p^{\beta}_{B^*_s}s^{\nu}_{l^-}s^{\sigma}_{l^+}\right)
 +m_lm_{B^*_s} \right. \nonumber\\
& & \left. \left(\mp 4m_{B^*_s}\sqrt{m^2_{B^*_s}-4m^2_l}+i\varepsilon_{\alpha\beta\nu\sigma}p^{\alpha}_{l^-}p^{\beta}_{B^*_s}s^{\nu}_{l^-}s^{\sigma}_{l^+}-i\varepsilon_{\beta\gamma\nu\sigma}p^{\beta}_{B^*_s}p^{\gamma}_{l^+}s^{\nu}_{l^-}s^{\sigma}_{l^+} \right)\right\rbrace \left. +\frac{2}{3}\left(m^2_{B^*_s}-4m^2_l\right)\vert C_{10}\vert^2 \right].
\label{gama2}
\end{eqnarray}
\end{widetext}
In eqs.~(\ref{gama1}) and (\ref{gama2}), we have used the abbreviations $\mathcal{N}=\frac{\alpha^2_{em}G^2_F}{128\pi^3}\vert V_{tb}V^*_{ts}\vert^2f^2_{B^*_s}\sqrt{m^2_{B^*_s}-4m^2_l}$, $\mathcal{C}=\left(C_9^{eff}+\frac{2 m_b f_{B_s^*}^T}{m_{B_s^*} f_{B_s^*}}C_7^{eff}\right)$.
Using eqs.~(\ref{alp}),(\ref{gama1}) and (\ref{gama2}), we get the lepton polarization asymmetry to be 
\begin{widetext}
\begin{equation}
\mathcal{A}_{LP}^{\pm}\vert_{SM}= \mp \frac{ 2\sqrt{1-4 m_{l}^2/m^2_{B_s^*}}~Re\left[\left(C_9^{eff}+\frac{2 m_b f_{B_s^*}^T}{m_{B_s^*} f_{B_s^*}}C_7^{eff} \right) C_{10}^*\right]}{\left(1 + 2 m_{l}^2/m_{B_s^*}^2 \right) \left|C_9^{eff}+\frac{2 m_b f_{B_s^*}^T}{m_{B_s^*} f_{B_s^*}}C_7^{eff}\right|^2 + \left(1 -4 m_{l}^2/m_{B_s^*}^2\right) \left|C_{10}\right |^2}.
\label{eqSM}
\end{equation}
\end{widetext}

%%%%%%%%%%%%%%%%%%%%%%%%%%
\subsection{Longitudinal polarization asymmetry in presence of NP}
%%%%%%%%%%%%%%%%%%%%%%
We now investigate the lepton polarization asymmetry in the presence of NP. As the NP solutions to the $b \rightarrow s l^+ l^-$ anomalies are in the form of $V$ and $A$ operators, we consider the addition of these NP operators to the SM effective Hamiltonian of $b \rightarrow s l^+ l^-$. Scalar and pseudo-scalar NP operators do not contribute to $B_s^*\rightarrow l^+l^-$ decay because $\langle0 \vert\bar{s} b \vert B_s^{*}(p_{B_s^*},\epsilon) \rangle = \langle0 \vert\bar{s} \gamma_5 b \vert B_s^{*}(p_{B_s^*},\epsilon) \rangle = 0$. The effective Hamiltonian now takes the form
\begin{align}
\mathcal{H}_{eff}(b \rightarrow s l^+ l^-) &= \mathcal{H}_{SM} + \mathcal{H}_{VA} ,
\end{align}
where $\mathcal{H}_{VA}$ is 
\begin{align} \nonumber
\mathcal{H}_{VA} &= \frac{\alpha_{em}\,G_F}{\sqrt{2} \pi} V_{ts}^* V_{tb} \bigg[C_9^{NP}(\overline{s} \gamma^{\mu} P_L b)(\overline{l} \gamma_{\mu} l) + C_{10}^{NP} (\overline{s} \gamma^{\mu} P_L b)(\overline{l} \gamma_{\mu} \gamma_{5} l) \bigg]. 
\end{align}
Here $C^{NP}_{9(10)}$ are the NP Wilson coefficients. Within this NP framework, the branching ratio and $\mathcal{A}_{LP}$ are obtained to be
\begin{widetext}
\begin{eqnarray}
\mathcal{B}(B^*_s\rightarrow l^+l^-) &=& \frac{\alpha^2_{em}G^2_Ff^2_{B^*_s}m^3_{B^*_s}\tau_{B^*_s}}{96\pi^3}\vert V_{ts}V^*_{tb}\vert^2\sqrt{1-4m_l^2/m^2_{B^*_s}}\left[\left(1+\frac{2m^2_l}{m^2_{B^*_s}}\right)\left\vert C_9^{eff}+\frac{2 m_b f_{B_s^*}^T}{m_{B_s^*} f_{B_s^*}}C_7^{eff}+C_9^{NP} \right\vert^2 \right. \nonumber\\
& &\left. +\left(1-\frac{4m^2_l}{m^2_{B^*_s}}\right)\vert C_{10}+ C_{10}^{NP}\vert^2\right],
\end{eqnarray}
\end{widetext}
\begin{widetext}
\begin{align}
\mathcal{A}_{LP}^{\pm}\vert_{NP}= \mp \frac{ 2\sqrt{1-4 m_{l}^2/m^2_{B_s^*}}~Re\left[\left(C_9^{eff}+\frac{2 m_b f_{B_s^*}^T}{m_{B_s^*} f_{B_s^*}}C_7^{eff}+C_9^{NP} \right) \left(C_{10}+C_{10}^{NP}\right)^*\right]}{\left(1 + 2 m_{l}^2/m_{B_s^*}^2 \right) \left|C_9^{eff}+\frac{2 m_b f_{B_s^*}^T}{m_{B_s^*} f_{B_s^*}}C_7^{eff} +C_9^{NP}\right|^2 + \left(1 - 4 m_{l}^2/m_{B_s^*}^2\right) \left|C_{10 }+C_{10}^{NP}\right|^2}.
\label{eqNP}
\end{align}
\end{widetext}

%%%%%%%%%%%%%%%%%%%%%%%
\section{Results and Discussion}
\subsection{ $\mathcal{A}_{LP}(\mu)$ with NP solutions}

In this section we first calculate $\mathcal{A}_{LP}(\mu)$ for the $B_s^* \rightarrow \mu^+ \mu^-$ decay. The numerical inputs used for this calculation are listed in table~\ref{tab1}. The SM prediction is \begin{equation}
\mathcal{A}^{+}_{LP}(\mu)\vert_{SM}=-\mathcal{A}^{-}_{LP}(\mu)\vert_{SM}= 0.9955\pm 0.0003.
\end{equation}The uncertainty in this prediction (about $0.03\%$) is much smaller than the uncertainty in the decay constants (about $2\%$), making it theoretically clean.
%%%%%%%%%%%%%%%%%%%%%%%%%%%%
\begin{table}[h]
	\centering 
	\begin{tabular}{|l|l|} \hline\hline	
		Parameter   &  Value  \\
		\hline
		$m_b$ & $4.18\pm 0.03$ GeV~\cite{Patrignani:2016xqp}  \\
		
		$m_{B^*_s}$  & $5415.4^{+1.8}_{-1.5}$ MeV~\cite{Patrignani:2016xqp} \\
				$f_{B^*_s}/f_{B_s}$ &  $0.953\pm 0.023$~\cite{Colquhoun:2015oha} \\
		
		$f^T_{B^*_s}/f_{B_s}$ &  $0.95$~\cite{Grinstein1509} \\
		\hline\hline
	\end{tabular}
	\caption{Numerical inputs used in our calculations.}
	\label{tab1}
\end{table}

Among the two NP solutions which can account for all the $b \rightarrow s \mu^+ \mu^-$ anomalies~\cite{Capdevila:2017bsm,AKAPRD96}, only $C_9^{NP}(\mu\mu)$ is non-zero for the first solution whereas $C_9^{NP}(\mu\mu)$ and $C_{10}^{NP}(\mu\mu)$ are equal and opposite for the second solution. In table~\ref{tab2} we have listed the NP WCs of these solutions along with the predictions of $\mathcal{A}_{LP}^{\pm}(\mu)$ for them.

%%%%%%%%%%%%%%%%%%%%%%%%%%%%%%%%%%%%%%
\begin{table*}[htbp]
	\centering 
	\tabcolsep 3pt
	\begin{tabular}{|c|c|c|c|}
		\hline\hline
		NP type  &  NP WCs & $\mathcal{B}(B^*_s\rightarrow \mu^+\mu^-)$ & $\mathcal{A}^+_{LP}(\mu)=-\mathcal{A}^-_{LP}(\mu)$  \\
		\hline
		SM & 0 & $(1.10\pm 0.60)\times 10^{-11}$ & $0.9955 \pm 0.0003$ \\ \hline
		(I) $C_9^{NP}(\mu\mu)$~~~~~~~~~~~~~ & $-1.25\pm 0.19$ & $(0.83\pm 0.45)\times 10^{-11}$ & $0.8877 \pm 0.0312$\\
		
		(II) $C_9^{NP}(\mu\mu)=-C_{10}^{NP}(\mu\mu)$  & $-0.68 \pm0.12$ & $(0.79\pm 0.43)\times 10^{-11}$  &$0.9936 \pm0.0057$ \\
		\hline\hline
	\end{tabular}
	\caption{ New physics predictions of branching ratio and $\mathcal{A}_{LP}(\mu)$ for $B_s^* \rightarrow \mu^+ \mu^-$ decay with real NP WCs.}
	\label{tab2}

\end{table*}
%%%%%%%%%%%%%%%%%%%%%% 
\begin{table*}[htbp]
	\centering 
	\tabcolsep 2pt
	\begin{tabular}{|c|c|c|c|}
		\hline\hline
		NP Type   &  [Re(WC), Im(WC)] & $\mathcal{B}(B^*_s\rightarrow \mu^+\mu^-)$ & $\mathcal{A}^+_{LP}(\mu)=-\mathcal{A}^-_{LP}(\mu)$  \\
		\hline
		(I) $C_9^{NP}(\mu\mu)$ & $[(-1.1\pm 0.2), (0.0\pm 0.9)]$ & $(0.85\pm 0.27)\times 10^{-11}$ & $0.91 \pm 0.13$\\
\hline		
(II) $C_9^{NP}(\mu\mu)=-C_{10}^{NP}(\mu\mu)$  & (A) $[(-0.8\pm 0.3), (1.2\pm 0.7)]$ & $(0.80\pm 0.27)\times 10^{-11}$  &$0.99 \pm0.02$ \\
		& (B) $[(-0.8\pm 0.3), (-1.2\pm 0.8)]$ & $(0.80\pm 0.28)\times 10^{-11}$ & $0.99\pm 0.11$\\
		\hline\hline
	\end{tabular}
	\caption{New physics predictions of branching ratio and $\mathcal{A}_{LP}(\mu)$ for $B_s^* \rightarrow \mu^+ \mu^-$ decay with complex NP WCs. The NP WCs are taken from ref.~\cite{AKAPRD96:015034}}
	\label{tab3}
\end{table*}

From this table it is obvious that the prediction of $\mathcal{A}_{LP}(\mu)$ for the first solution deviates from the SM at the level of $3.4\sigma$ whereas, for the second solution, it is the same as that of the SM. Hence any large deviation in this asymmetry can only be due to the first NP solution. We also provide the predictions for $\mathcal{B}(B^*_s\rightarrow \mu^+\mu^-)$ in table~\ref{tab2}. It is clear that neither of the two solutions can be distinguished from each other or from the SM via the branching ratio.

In the discussion above, the NP WCs are assumed to be real. If these WCs are complex, they can lead to various CP asymmetries in $B\rightarrow (K,K^*)\mu^+\mu^-$ decays~\cite{Alok:2011gv}. These asymmetries can distinguish between the two NP solutions. In ref.~\cite{AKAPRD96:015034}, it was assumed that $C_9^{NP}(\mu\mu)$ and $C_{10}^{NP}(\mu\mu)$ are complex and a fit to all the $b\rightarrow s\mu^+\mu^-$ data was performed. The resulting values of NP WCs from this fit are given in table~\ref{tab3}. The predictions for $\mathcal{B}(B^*_s\rightarrow \mu^+\mu^-)$ and $\mathcal{A}_{LP}(\mu)$ are also given in this table. Because of the large uncertainties, neither of these two observables can distinguish between the two NP solutions. However, it is possible to make a distinction based on the CP asymmetries mentioned above~\cite{AKAPRD96:015034}.

\begin{figure*}[htbp]
	\begin{tabular}{cc}
		\includegraphics[width=86mm]{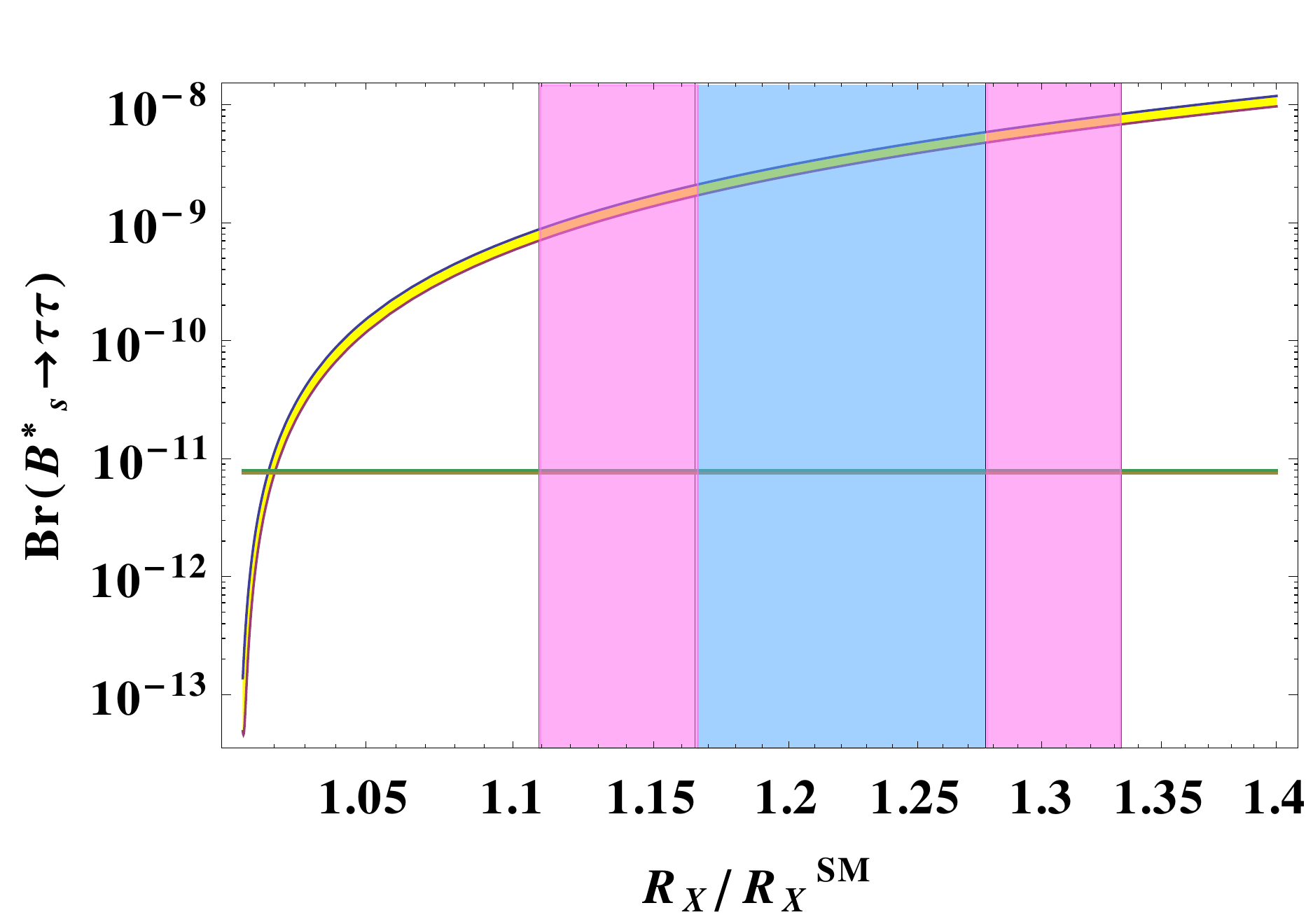}&
		\includegraphics[width=87mm]{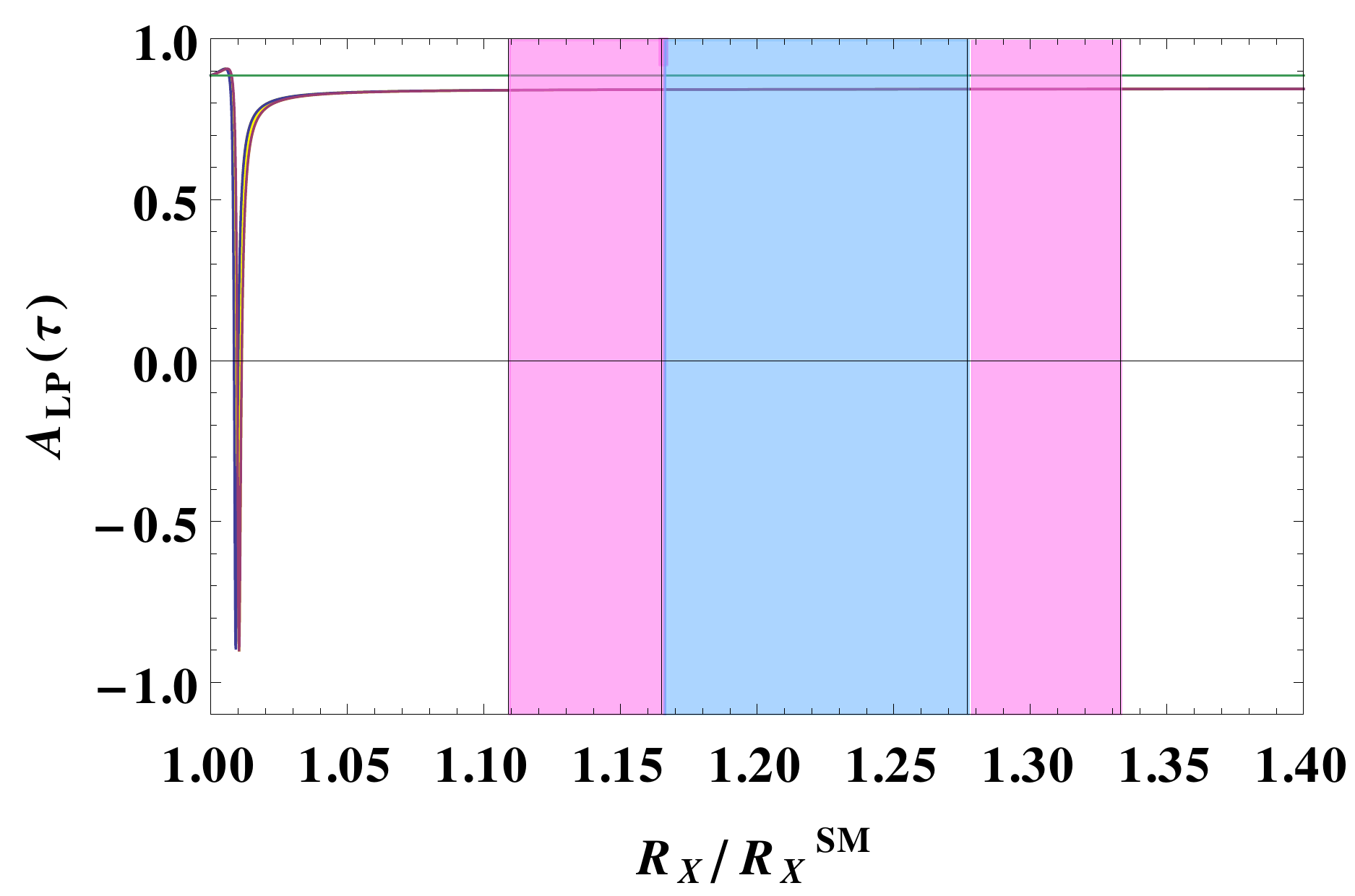}\\
	\end{tabular}
	\caption{Left and right panels correspond to $\mathcal{B}(B^*_s\rightarrow \tau^+\tau^-)$ and $\mathcal{A}_{LP}(\tau)$ respectively. In both panels the yellow band represents $1\sigma$ range of these observables. The $1\sigma$ and $2\sigma$ ranges of $R_X/R^{SM}_X$ are indicated by blue and pink bands respectively. The green horizontal line corresponds to the SM value.} 
	\label{fig1}
\end{figure*}

\subsection{Effect of NP in $B_s^* \rightarrow \tau^+ \tau^-$ }

 As mentioned in the introduction, anomalies are also observed in the $b\rightarrow c\tau\bar{\nu}$ transitions. An NP model, which can account for these anomalies, is likely to contain NP amplitude for $b\rightarrow s\tau^+\tau^-$ transition also. Hence the branching ratio of $B^*_s\rightarrow \tau^+\tau^-$ and $\tau$ longitudinal polarization asymmetry $\mathcal{A}_{LP}(\tau)$ will contain signatures of such NP. In the SM, the predictions for these quantities are 
 \begin{eqnarray}
& \mathcal{B}(B^*_s\rightarrow \tau^+\tau^-) = (6.87 \pm 4.23) \times 10^{-12},\\
 & \mathcal{A}^+_{LP}(\tau)|_{SM}=-\mathcal{A}^{-}_{LP}(\tau)|_{SM} = 0.8860\pm 0.0006.
 \end{eqnarray}
The authors of ref.~\cite{Capdevila:2017iqn} constructed a model of NP which accounts for the anomalies in $b\rightarrow c\tau \bar{\nu}$. This model contains tree level FCNC terms for $b\rightarrow s\,\tau^+\,\tau^-$ but not for $b\rightarrow s l^+ l^-$ ($l=e,\mu$). Therefore, the WCs $C^{NP}_9(\mu\mu)$ and $C^{NP}_{10}(\mu\mu)$ have no relation to the WCs $C^{NP}_9(\tau\tau)$ and $C^{NP}_{10}(\tau\tau)$. The amplitude for $b\rightarrow s\mu^+\mu^-$ transition remains small enough that the constraints from $R_K$ and $R_{K^*}$ are satisfied. The WCs for the $b\rightarrow s\tau^+\tau^-$ transition have the form
\begin{eqnarray}
C_{9}(\tau\tau) &=& C^{SM}_{9} - C^{NP}(\tau\tau), \nonumber \\ 
C_{10}(\tau\tau) &=& C^{SM}_{10} + C^{NP}(\tau\tau),
\label{np}
\end{eqnarray}
in this model, where 
\begin{equation}
C^{NP}(\tau\tau) = \frac{2\pi}{\alpha}\frac{V_{cb}}{V_{tb}V^*_{ts}}\left(\sqrt{\frac{R_X}{R^{SM}_X}}-1\right).
\end{equation}
The ratio $R_X/R^{SM}_X$ is the weighted average of current experimental values of $R_D$, $R_{D^*}$ and $R_{J/\psi}$. From the current measurements of these quantities, we estimate this ratio to be $\simeq 1.22\pm 0.06$. This, in turn, leads to $C^{NP}(\tau\tau)\sim \mathcal{O}(100)$. Thus the NP contribution completely dominates the WCs and leads to greatly enhanced branching ratios for various $B$/$B_s$ meson decays involving $b\rightarrow s\,\tau^+\,\tau^-$ transition~\cite{Capdevila:2017iqn}. 
   
We calculate $\mathcal{B}(B^*_s\rightarrow \tau^+\tau^-)$ and $\mathcal{A}_{LP}(\tau)$ as a function of $R_X/R^{SM}_X$.
The plot of $\mathcal{B}(B^*_s\rightarrow \tau^+\tau^-)$ $vs.$ $R_X/R^{SM}_X$ is shown in left panel of fig.~\ref{fig1}. We note, from this plot, that  $\mathcal{B}(B_s^* \rightarrow \tau^+ \tau^-)$ can be enhanced up to $~10^{-9}$ which is about three orders of magnitude larger than the SM prediction. 
The plot of $\mathcal{A}_{LP}(\tau)$ $vs.$ $R_X/R^{SM}_X$ is shown in the right panel of fig.~\ref{fig1}. It can be seen that $\mathcal{A}_{LP}(\tau)$ is suppressed by about $5\%$ in comparison to its SM value.

The recent data on $R_{D^{(*)}}$ show less tension with the SM which leads to smaller values of $R_X/R_X^{SM}$. As long as this ratio is greater than $ 1.05$, the branching ratio of $B^*_s\rightarrow \tau^+\tau^-$ is enhanced by an order of magnitude at least. When $R_X/R^{SM}_X \sim 1.01$, $\mathcal{A}_{LP}(\tau)$ exhibits some very interesting behaviour. In this case, the tree level FCNC NP contribution is similar in magnitude to the SM contribution (which occurs only at loop level). Due to the interference between these two amplitudes, $\mathcal{A}_{LP}(\tau)$ changes sign and becomes almost ($-1$). Hence a measurement of this asymmetry provides an effective tool for the discovery of tree level FCNC amplitudes of this model~\cite{Capdevila:2017iqn} when their magnitude becomes quite small.

%%%%%%%%%%%%%%%%%%%%%%%%%%%%
\section{Conclusions}
%%%%%%%%%%%%%%%%%%%%%%%%%%%%
There are several measurements in the decays induced by the quark level transition $b\rightarrow s l^+l^-$ which do not agree with their SM predictions. All these discrepancies can be explained by considering NP only in $b\rightarrow s \mu^+\mu^-$ transition. These NP operators are required to have $V$ and/or $A$ form to account for the fact that $R_K$ and $R_{K^*}$ are less than 1. A global analysis of  all the measurements in $b\rightarrow s l^+l^-$ sector leads to only two NP solutions. The first solution has $C_9^{NP}(\mu\mu)<0$  and the second has $C_9^{NP}(\mu\mu)=-C_{10}^{NP}(\mu\mu)<0$. In this work we consider the ability of the muon longitudinal
polarization asymmetry in $B_s^* \rightarrow \mu^+ \mu^-$ decay to distinguish
between these two solutions. This observable is theoretically clean because it has only a very mild dependence on the decay constants. For the case of real NP WCs, we show that this asymmetry has the same value as the SM case for the second solution but is smaller by $11\%$ for the first solution. Hence, a measurement of this asymmetry to $10\%$ accuracy can distinguish between these two solutions. But for the complex NP WCs,  the discrimination power is lost because of the large theoretical uncertainties.

Further, we study the impact of the anomalies in $b\rightarrow c\tau\bar{\nu}$ transitions on the branching ratio of $B_s^* \rightarrow \tau^+ \tau^-$ and $\mathcal{A}_{LP}(\tau)$. In ref.~\cite{Capdevila:2017iqn}, a model was constructed where tree level NP leads to both $b\rightarrow s\tau^+\tau^-$ and $b\rightarrow c\tau\bar{\nu}$ with moderately large NP couplings. Within this NP model, we find that the present data in $R_{D^{(*)},J/\psi}$ sector imply about three orders of magnitude enhancement in the branching ratio of $B_s^* \rightarrow \tau^+ \tau^-$ and a $5\%$ suppression  in $\mathcal{A}_{LP}(\tau)$ compared to their SM predictions. We also show that $\mathcal{A}_{LP}(\tau)$ undergoes drastic changes when the NP amplitude is similar in magnitude to the SM amplitude.

To measure $\mathcal{A}_{LP}(\mu)$ or $\mathcal{A}_{LP}(\tau)$ in experiments, the final state leptons have to decay into secondary particles. But for muon, the measurement would be quite difficult as it does not decay within the detector. In the case of $\mathcal{A}_{LP}(\tau)$, it may be possible for LHCb to reconstruct $\tau$ where the $\tau$ decays into multiple hadrons. This technique has been already used to identify the $\tau$ leptons in $B\rightarrow D^*\tau\bar{\nu}$ decay. Therefore a precise reconstruction from the decay products of the $\tau$ is necessary to measure the $\tau$ longitudinal polarization asymmetry in $B_s^* \rightarrow \tau^+ \tau^-$. 

%%%%%%%%%%%%%%%%%%%%
\section*{Acknowledgements}
%%%%%%%%%%%%%%%%%%%%%%%%
We thank S. Uma Sankar and Ashutosh Kumar Alok for careful reading of the manuscript and their valuable comments and suggestions.
%%%%%%%%%%%%%%%%%%%%%%%%%%%%%%%%%%%%

\end{document}